# Controlling quantum interference between virtual and dipole two-photon optical excitation pathways using phase-shaped laser pulses

J. Lahiri[1], S. H. Yuwono[1], I. Magoulas[1], M. Moemeni[1], B. Borhan[1], G. J. Blanchard[1], P. Piecuch[1,2], M. Dantus[1,2,*]

[1] Department of Chemistry, Michigan State University, East Lansing, MI 48824, USA

[2] Department of Physics and Astronomy, Michigan State University, East Lansing, MI 48824, USA

* Corresponding author. Email: dantus@chemistry.msu.edu

## ABSTRACT

Two-photon excitation (TPE) proceeds via a "virtual" pathway, which depends on the accessibility of one or more intermediate states, and, in the case of non-centrosymmetric molecules, an additional "dipole" pathway involving the off-resonance dipole-allowed one-photon transitions and the change in the permanent dipole moment between the initial and final states. Here, we control the quantum interference between these two optical excitation pathways by using phase-shaped femtosecond laser pulses. We find enhancements by a factor of up to 1.75 in the two-photon-excited fluorescence of the photobase **FR0**-SB in methanol after taking into account the longer pulse duration of the shaped laser pulses. Simulations taking into account the different responses of the virtual and dipole pathways to external fields and the effect of pulse shaping on two-photon transitions are found to be in good agreement with our experimental measurements. The observed quantum control of TPE in condensed phase may lead to enhanced signal at a lower intensity in two-photon microscopy, multiphoton-excited photoreagents, and novel spectroscopic techniques that are sensitive to the magnitude of the contributions from the virtual and dipole pathways to multiphoton excitations.

# INTRODUCTION

Control of two-photon transitions using shaped laser pulses in atomic[1] and condensed phase[2–4] systems has primarily been accomplished by taking advantage of the direct relationship between the second-order power spectrum (SOPS) of the field and the spectral phase of the pulse. In fact, this dependence has enabled selective two-photon excitation (TPE) and the measurement of TPE spectra.[5–10] Here, we address the quantum interference between the virtual and dipole optical pathways that connect the ground and excited electronic states of a molecule during TPE. In particular, we consider the case of a non-centrosymmetric molecule for which, in addition to the sum over intermediate "virtual" states, one is able to identify a dipole contribution which becomes significant for species with large change in the permanent dipole moment upon photoexcitation.

In their theoretical investigations, Meath and coworkers proposed how the phase difference between a pair of monochromatic pulses could be used to control two-photon transitions in isolated diatomic molecules with permanent dipoles by taking advantage of the different responses of the virtual and dipole pathways to the electric fields.[11,12] In a related experiment, the TPE of Rb atoms through an intermediate state was found to be favored by phase- or amplitude-shaped pulses rather than by transform-limited (TL) pulses.[13] It was observed that by removing certain portions of the pulse spectrum, which decreased the peak intensity by a factor of 40, the TPE rate doubled relative to TL pulses.[13] The observed control was due to differences in the optical response to the field by the resonant and non-resonant excitation pathways. While contributions from both virtual and dipole pathways have been implicated in the magnitude of the two-photon absorption cross section,[14,15] their control via phase-shaped pulses has not been reported. For a given molecule, the inherent molecular phases associated with the virtual and dipole pathways are fixed. However, the goal of the present work is to use phase-shaped pulses designed to drive the virtual and dipole pathways with an independently adjustable phase, so that one is able to control the probability of TPE.

Here, we control the TPE pathways of the fluorescent Schiff photobase (E)-7-((butylimino)methyl)-N,N-diethyl-9,9-dimethyl-9H-fluoren-2-amine (**FR0**-SB, Scheme 1), in solution.[16–18] This compound, in addition to exhibiting an increase in $pK_a$ by 14 units upon photoexcitation, was recently found to become up to 62% more reactive when the first excited singlet (S$_1$) state was populated via TPE instead of single-photon excitation.[19] This is a consequence of the very large differences between the ground- and excited-state dipole moments $\Delta\mu_{10} = \mu_1 - \mu_0$ and the large values of the corresponding transition dipole moments $\mu_{10}$ (for example, $\Delta\mu_{10} = 15.6$ D and $\mu_{10} = 9.6$ D for **FR0**-SB in methanol).[19] In this study, we develop the theoretical basis for the effect of shaped pulses, which are far from one-photon resonance with the electronic transition to S$_1$ but on-resonance with TPE to the S$_1$ state, on the virtual and dipole pathway contributions to TPE of molecules with large permanent dipole moments. We then present experimental data showing coherent control of the virtual and dipole contributions to TPE. In particular, we find that for certain shaped pulses the ratio between TPE fluorescence and second harmonic signal can be enhanced by a factor of 1.75, in good agreement with simulations. Our study demonstrates that it is possible to use phase-shaped pulses to control the dipole and virtual contributions to TPE in larger polyatomic molecules in solution.

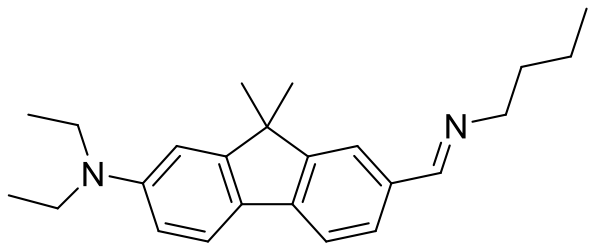

**Scheme 1**. Molecular structure of the **FR0**-SB super photobase.

## THEORY

Since the introduction of the idea of two-photon processes by Maria Göppert-Mayer in 1931[20] (cf., also, refs 21 and 22), perturbation theory has been used to examine their probability and dependence on molecular symmetry. According to the second-order perturbation theory, the $0 \to f$ two-photon absorption cross section, with 0 and $f$ designating the initial and final states, which in our case are the electronically bound ground (S$_0$) and excited (S$_1$) states of **FR0**-SB, respectively, is given by the Göppert-Mayer sum over virtual states

$$\sigma_{f0}^{(2)}(\omega/2) = A \left| \sum_{v} \frac{\mu_{fv}\mu_{v0}}{\omega_{v0} - \omega/2 + i\Gamma_{v}(\omega/2)} \right|^2 g_{M2}(\omega), \tag{1}$$

where $\omega$ is the resonance excitation frequency, $A$ is a collection of constants, $v$ are the intermediate states, $\mu_{fv}$ and $\mu_{v0}$ are the transition dipole moments corresponding to the $v \to f$ and $0 \to v$ excitations, respectively, $\omega_{v0}$ is the $0 \to v$ transition frequency, $i\Gamma_v(\omega/2)$ is a damping factor associated with the intermediate state $v$, which is inversely proportional to the lifetime of $v$, and $g_{M2}(\omega)$ is the TPE line shape function. Note that in presenting eq 1, we have performed an isotropic averaging over the directions of the transition dipole moment vectors $\boldsymbol{\mu}_{fv}$ and $\boldsymbol{\mu}_{v0}$. For non-centrosymmetric molecules, one can separate the $0 \to f$ dipole contribution from the sum over states expression, eq 1, to obtain[14,19,23,24]

$$\sigma_{f0}^{(2)}(\omega/2) = A \left| \sum_{v \neq 0,f} \frac{\mu_{fv}\mu_{v0}}{\omega_{v0} - \omega/2 + i\Gamma_v(\omega/2)} + \frac{\mu_{f0}\Delta\mu_{f0}}{\omega/2} \right|^2 g_{M2}(\omega), \quad (2)$$

where $\Delta\mu_{f0} = \mu_f - \mu_0$ is the difference between the dipole moments characterizing the final and initial states. To do this, we isolate the $v = 0$ and $v = f$ terms in the summation over $v$ in eq 1, $\mu_{f0}\mu_{00}/[\omega_{00} - \omega/2 + i\Gamma_0(\omega/2)]$ and $\mu_{ff}\mu_{f0}/[\omega_{f0} - \omega/2 + i\Gamma_f(\omega/2)]$, respectively, which involve off-resonance dipole-allowed one-photon $0 \to f$ transitions, recognize that $\omega_{00} = 0$, replace $\omega_{f0}$ by the resonance frequency $\omega$, and take advantage of the fact that the initial and final states are stationary, so that the damping factors $i\Gamma_0(\omega/2)$ and $i\Gamma_f(\omega/2)$ can be assumed to be zero. The resulting $\mu_{f0}\mu_{00}/(-\omega/2)$ and $\mu_{ff}\mu_{f0}/(\omega/2)$ contributions can be recombined into the expression $\mu_{f0}(\mu_{ff} - \mu_{00})/(\omega/2)$, which is equivalent to the second term on the right-hand side of eq 2. The first term on the right-hand side of eq 2, which involves the intermediate states $v$ between the initial and final states, is usually called the virtual pathway, while the second term, which depends on the transition dipole moment $\mu_{f0}$ and the difference between the ground- and excited-state permanent dipole moments $\Delta\mu_{f0}$, is commonly referred to as the dipole pathway.[11] It is worth mentioning that while the cross-section contributions from both pathways are proportional to the intensity of light squared, the dipole pathway, which is, in part, driven by off-resonance one-photon transitions to a stationary state $f$, has a weaker dependence on the pulse duration compared to its virtual counterpart.

The introduction of femtosecond laser pulses has greatly contributed to the development of TPE techniques. This is because broadband femtosecond pulses enable contributions from a very large number of possible intermediate states, ensuring that some frequencies $\omega_{\nu 0}$ in eqs 1 and 2 are close to $\omega/2$, and the high peak intensity of such pulses reduces the need for the $i\Gamma_\nu(\omega/2)$ damping factors to become small to result in a measurable TPE signal. It should also be noted that non-centrosymmetric molecules, such as **FR0**-SB examined in this work, have a non-zero $\Delta\mu_{f0}$, so that both the virtual and dipole pathways present in eq 2 can contribute to TPE, further enhancing the two-photon absorption cross sections.[15]

The two pathways available for TPE have a phase difference that arises from the complex nature of the different matrix elements in eq 2. This phase difference may lead to quantum interference that could modulate the two-photon absorption cross section. Jagatap and Meath considered the competition between the virtual and dipole pathways and derived an expression for the orientationally averaged probability of TPE.[11] Based on their calculations for the LiH molecule, the quantum interference between the virtual and dipole pathways can be constructive or destructive. While TPE is fundamentally different than Young's double slit experiment or quantum control of molecular excitations, one might try to adopt a heuristic approach and use analogies with these phenomena to explain the possibility of modulating the probability for TPE through quantum interference of the virtual and dipole pathways. For example, if $\phi_1$ and $\phi_2$ designate phases associated with interfering waves in Young's double slit experiment, the probability of constructive interference $P$ depends on the phase difference according to $P \propto \left|e^{i\phi_1} + e^{i\phi_2}\right|^2 = 2 + 2\cos(\phi_1 - \phi_2)$. Similar interference can be observed in quantum control of molecular excitations, introduced by Brumer and Shapiro[25] and realized experimentally by Gordon and coworkers for one- versus three-photon transitions.[26] Gordon and coworkers' excitation pathways involved two laser sources with frequencies $3\omega$ and $\omega$. Assuming that each pathway contributes equally, the probability for reaching the final state is given by $P_{f0} \propto \left|e^{i\phi_3} + e^{i3\phi_1}\right|^2 = 2 + 2\cos(\phi_3 - 3\phi_1)$, where $\phi_3$ and $\phi_1$ are the phases of one- and three-photon excitations, respectively. Chen et al. have also examined quantum control of molecular excitation using two-photon vs. two-photon interference, where a molecule is irradiated with three interrelated frequencies $\omega_0$, $\omega_+$, and $\omega_-$, such that $2\omega_0 = \omega_+ + \omega_-$, and where the probability for reaching the final state is given by $P_{f0} \propto \left|e^{i2\phi_0} + e^{i(\phi_+ + \phi_-)}\right|^2 = 2 + 2\cos(2\phi_0 - \phi_+ - \phi_-)$ (see refs 27 and 28). The experimental realization of their idea was accomplished by broadband shaped femtosecond pulses, both in the gas phase[1,29,30] and in condensed phase systems.[2,3] Two-photon transitions involving virtual and dipole pathways should be controllable in a similar manner. Indeed, if we assume that none of the intermediate states $v$ in eq 2 are long lived and designate the average phase of the virtual pathway as $\phi_v$ and the phase associated with the dipole pathway as $\phi_d$, we can use arguments similar to those presented above to re-express the $0 \to f$ two-photon absorption cross section as

$$\sigma_{f0}^{(2)}(\omega/2) \propto \left| e^{i2\phi_v} + e^{i2\phi_d} \right|^2 = 2 + 2\cos(2\phi_v - 2\phi_d). \tag{3}$$

This shows that one should be able to modulate the probability for TPE through quantum interference of the two pathways.

We now describe the model which will allow us to examine how the interference between the virtual and dipole pathways contributing to TPE in non-centrosymmetric molecules, such as **FR0**-SB, is affected by broadband-shaped laser pulses. Two-photon transitions in the absence of long-lived intermediate states are driven by the shaped laser field, which induces a nonlinear polarization proportional to the square of the time-dependent electric field $E(t)$. Because pulse shapers operate in the frequency domain,[31] we start with the field in the frequency domain

$$E(\omega/2) = \sqrt{I(\omega/2)} e^{i\bar{\varphi}(\omega/2)}, \tag{4}$$

where $I(\omega/2)$ is the spectrum of the pulse and $\bar{\varphi}(\omega/2)$ is the spectral phase, both of which can be modified by the pulse shaper. Using the field in the frequency domain and the convolution theorem, the Fourier transform of $E^2(t)$ can be written as[3,32]

$$E^{(2)}(\omega) \propto \int_{-\infty}^{\infty} |E(\omega/2+\Omega)| |E(\omega/2-\Omega)| \exp\left\{i\left[\bar{\varphi}(\omega/2+\Omega) + \bar{\varphi}(\omega/2-\Omega)\right]\right\} d\Omega, \tag{5}$$

where one integrates over all the detuning frequencies from $\omega/2$ by a positive and negative amount $\Omega$, as the two-photon frequency $\omega$ can be attained by adding the negatively detuned frequencies to the corresponding positively detuned frequencies. Equation 5 is important for us, since one can use it to simulate the SOPS by determining the values of $\left|E^{(2)}(\omega)\right|^2$ in which the phase of the field at frequency $\omega$ is given by the sum inside the square brackets appearing in eq 5. The simulated SOPS can then be directly compared to the second harmonic generation (SHG) spectrum, which is also given by $\left|E^{(2)}(\omega)\right|^2$, provided that the SHG crystal is thin enough to phase match the entire bandwidth of the pulse. The dependence of SHG on spectral phase,[10] such as chirp and third-order dispersion, has been used by the Dantus group to develop a number of highly accurate pulse characterization and compression methods based on the principle of multiphoton intrapulse interference phase scan (MIIPS),[33–42] which is employed in this study as well.

For simulating and carrying out the control experiments reported in this work, we used well-defined spectral functions to ensure that the observed quantum control is not caused by tuning the frequency of the TPE field. The form of the spectral phase functions, which we adopted in this study and which is shown in Figure 1a, resembles a window having zero phase for all frequencies except for a region of given width in the center of the spectrum that has a constant phase which is scanned during the experiment. By inspecting eq 5, one can obtain the phase of the field at frequency $\omega$ as the sum of the phases that are symmetrically detuned around $\omega/2$. In the present study, we report results for the phase value scanned from $-2\pi$ to $2\pi$ in order to modulate the relative contributions from the dipole and virtual pathways, as further discussed below. When the phase has a value $n\pi$, with $n$ an even integer, the pulses are of the TL type and favor the virtual pathway. For phases where $n$ is an odd integer, the pulses are longer and favor the dipole pathway.

The expected effect of the phase scan on the SOPS, obtained as a contour plot of $\left|E^{(2)}(\omega)\right|^2$ calculated using eq 5 for 16.6 fs pulses, is shown in Figure 1b. In Figure 1c, we show the experimentally obtained SHG spectrum as a function of the 40 nm phase window being scanned from $-2\pi$ to $2\pi$. The excellent agreement between theory (Figure 1b) and experiment (Figure 1c) required careful pulse compression and a very thin SHG crystal, both described in the Experimental Details section below. The data in Figure 1b and c show areas where constructive (red color, when the phase window is 0 or $\pm 2\pi$) and partial destructive (blue color, when the phase window is $\pm\pi$) interferences take place.

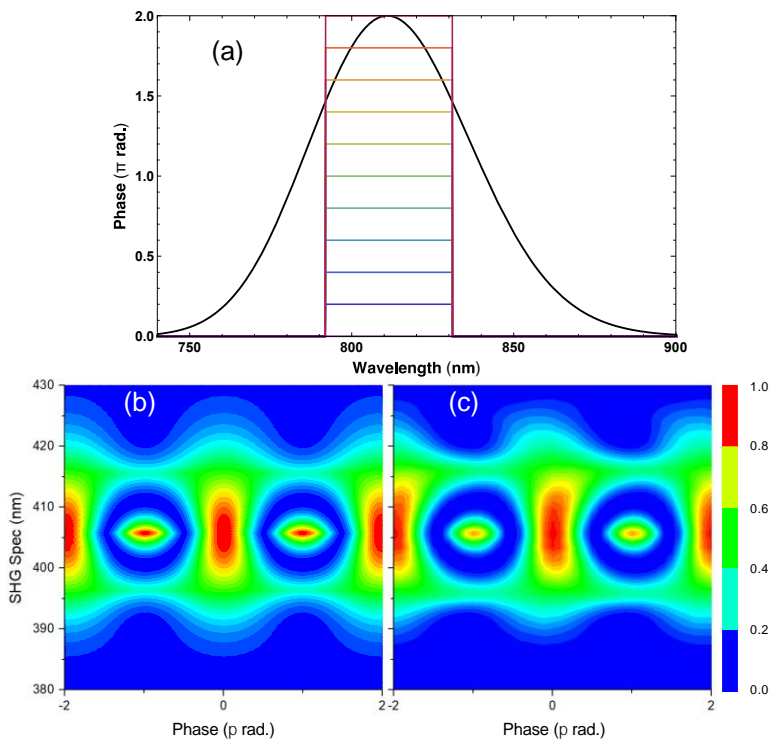

**Figure 1**. (a) A 40 nm wide spectral phase window centered on the spectrum of 16.6 fs Gaussian pulses at 811 nm resulting from the phase scanning from 0 to $2\pi$. (b) Calculated SOPS resulting from scanning the phase window from $-2\pi$ to $2\pi$. (c) Experimental SHG spectrum as a function of the same phase window scan as used in (b). In panels (b) and (c), areas where constructive and partial destructive interferences take place are colored red and blue, respectively.

The efficiency of TPE depends not only on the phase of the pulse, but also on the overlap between the SOPS and the two-photon absorption spectrum of the chromophore. Therefore, we need to make sure that the latter factor stays constant, given that our interest is to control the interference between the virtual and dipole pathways in TPE. As mentioned above, the chosen phase window function preserves the central wavelength of the SOPS, ensuring that changes in excitation probability are minimized (Figure 1b and c). We provide further verification of this in Figure 2. Specifically, in Figure 2a we present the SOPS data for shaped pulses using a 40 nm wide phase window with different amplitudes, plotted together with the TPE spectrum of **FR0**-SB in methanol using the data points reported in ref 19. As demonstrated in Figure 2a, the SOPS remains centered at 405.5 nm as the amplitude of the phase window changes. In Figure 2b we show the SHG intensity, the simulated two-photon excited fluorescence (TPEF) corresponding to the overlap integral between the SOPS and the TPE spectrum of **FR0**-SB, and the ratio between these two quantities as functions of the phase of the 40 nm window. We find that the overlap between the SOPS and the two-photon absorption spectrum of the chromophore differs only very little from the SHG signal, as shown in Figure 2b. Hence, the ratio between these two quantities remains close to 1, with variations of less than 3% for all phase amplitudes. Although not shown in Figure 2b, the same is observed for the other window widths used in this work.

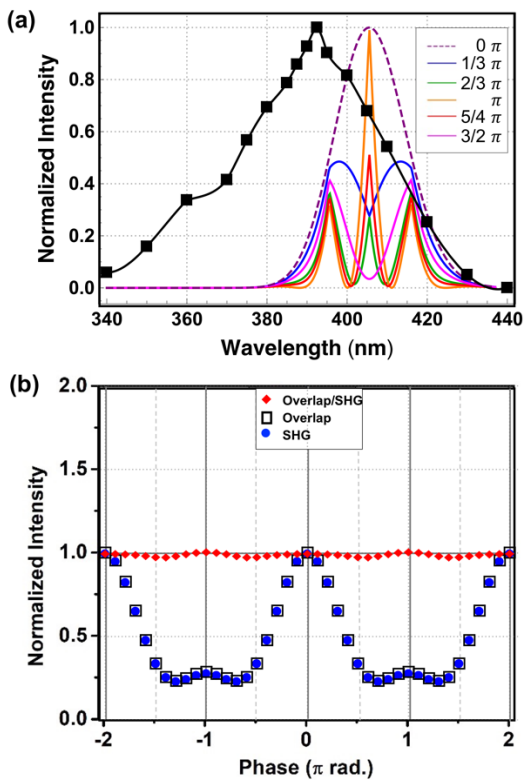

**Figure 2**. (a) The TPE spectrum of **FR0**-SB in methanol (black squares), based on the data points reported in ref 19, along with the SOPS for the 40 nm wide phase window at selected amplitudes. Note that as the phase amplitude changes the SOPS remains centered at 405.5 nm. (b) The results of simulations of the TL-normalized overlap integral between the SOPS and the TPE spectrum of **FR0**-SB in methanol (black open squares), the TL-normalized SHG signal (blue circles), and the ratio of the two signals (red diamonds) as functions of the spectral phase of the 40 nm window.

Having demonstrated the validity of eq 5, we transform eq 2, with the help of eqs 4 and 5, into a formula that one can use to predict how two-photon transitions in non-centrosymmetric molecules are affected by broadband-shaped laser pulses. Given that the coherent superpositions of tens or hundreds of intermediate states $v$ in eq 2, which are expected to have small Frank−Condon factors, are short lived, we assume that the virtual pathway behaves as a parametric process. Thus, we describe the virtual pathway using eq 5, which is similar to the approximation adopted by Silberberg and coworkers in simulating the non-resonant TPE path.[13] As implied by eq 2, the dipole pathway contains the action of the field on the transition dipole moment $\mu_{f0}$ and on the difference between the ground- and excited-state permanent dipoles $\Delta\mu_{f0}$, which include the aforementioned phase lag, designated here as $\phi$. Thus, given the fact that the dipole-allowed one-photon transitions between the initial and final stationary states associated with $\mu_{f0}$ are not significantly affected by pulse duration, we represent the dipole pathway contribution to eq 2 by multiplying the square of $E(\omega/2)$, as defined by eq 4, by the phase lag term $e^{i\phi}$, while incorporating the orientationally averaged $\mu_{f0}$ and $\Delta\mu_{f0}$ values in an empirical constant $a$ that describes the relative contribution of the two pathways. The resulting expression for the probability of TPE, which allows us to simulate the effect of shaped laser pulses on two-photon transitions in non-centrosymmetric molecules, is as follows:

$$P^{(2)}_{f0}(\omega/2) \propto \left| E^{(2)}(\omega) + \frac{aE^2(\omega/2)e^{i\phi}}{\omega/2} \right|^2 g_{M2}(\omega). \tag{6}$$

In transitioning from eq 2 to eq 6, we adopted a simplifying assumption where the contributions of all intermediate states $v$ to the TPE absorption cross section, including vibronic components, are lumped into a single electronic state. While we realize the limitations of this assumption, expanding our analysis to explicitly include vibrational modes that may be accessed by the broadband pulse and the effects of homogeneous and inhomogeneous broadening caused by the solvent is outside of our present modeling capabilities. Note that because our experiments are carried out with broadband femtosecond laser pulses, the determination of the $E^{(2)}(\omega)$ contribution to eq 6 for a given spectral phase function $\bar{\varphi}(\omega/2)$ requires the integration over all $\Omega$ values in eq 5.

In the experiments reported in this work, we measured the frequency integrated TPEF signal obtained for shaped pulses, $S^{\text{TPEF}}_{\text{shaped}}$, relative to that obtained for TL pulses, $S^{\text{TPEF}}_{\text{TL}}$. Based on the considerations described above, one should be able to model the ratio of the frequency integrated TPEF signals obtained using shaped and TL pulses using the following expression:

$$S^{\text{TPEF}}_{\text{shaped}} \Big/ S^{\text{TPEF}}_{\text{TL}} = \frac{\int_{-\infty}^{\infty} \left| E^{(2)}(\omega) + aE^2(\omega/2)e^{i\phi}/(\omega/2) \right|^2_{\text{shaped}} d(\omega/2)}{\int_{-\infty}^{\infty} \left| E^{(2)}(\omega) + aE^2(\omega/2)e^{i\phi}/(\omega/2) \right|^2_{\text{TL}} d(\omega/2)}. \tag{7}$$

In writing eq 7, we assumed that the TPE line shape $g_{M2}(\omega)$ in eq 6 is a slowly varying function of $\omega$, so that one can cancel it out. Indeed, when we included the TPE line shape function $g_{M2}(\omega)$ in the determination of the $S_{shaped}^{TPEF} / S_{TL}^{TPEF}$ values using the two-photon absorption cross section of **FR0**-SB in methanol from ref 19, its effect on the simulated signal was very small. We used eq 7, with $E^{(2)}(\omega)$ and $E(\omega/2)$ determined using eqs 5 and 4, respectively, in our simulations, comparing the resulting $S_{shaped}^{TPEF} / S_{TL}^{TPEF}$ values obtained at the different phases and window widths with the experimental data. In determining the spectrum of the pulse entering eq 4, we used the expression $I(\omega/2) = e^{-[\tau_f(\omega/2-\omega_0)/g]^2}$, where $\tau_f$ is the pulse duration full-width at half maximum (FWHM) of the Gaussian pulses, $g = 2\sqrt{\log 2}$, and $\omega_0$ is the center frequency of the laser. In the simulations shown in this work, we used $\tau_f = 16.6$ fs and $\omega_0 = 2\pi c / \lambda$, with c being the speed of light and $\lambda = 811$ nm, and the phase window was centered at the center frequency of the laser pulses. The phase lag associated with the dipole pathway that best reproduced the experimental data was $\phi = 0.46\pi$, and the value of the empirical constant $a$ in eq 7 that worked best was 0.075. In simulating the $S_{shaped}^{TPEF} / S_{TL}^{TPEF}$ ratio as a function of the phase window, we varied the window phase from $-2\pi$ to $2\pi$ and the window width from 10 nm to 50 nm. We also simulated the ratio of the frequency integrated SHG signals obtained for shaped and TL pulses as a function of the phase window using the formula

$$S_{shaped}^{SHG} \Big/ S_{TL}^{SHG} = \frac{\int_{-\infty}^{\infty} \left|E^{(2)}(\omega)\right|^2_{shaped} d(\omega/2)}{\int_{-\infty}^{\infty} \left|E^{(2)}(\omega)\right|^2_{TL} d(\omega/2)}, \quad (8)$$

where $E^{(2)}(\omega)$ was calculated using eq 5 and the window phase and width were varied in the same way as for $S_{shaped}^{TPEF} / S_{TL}^{TPEF}$, comparing the resulting $S_{shaped}^{SHG} / S_{TL}^{SHG}$ with experiment. Finally, we calculated the ratios of the TL-normalized TPEF and SHG signals obtained using eqs 7 and 8, respectively, and compared them to their experimentally determined counterparts. All of the simulations reported in this work were carried out by programming eqs 4–8 using Wolfram Mathematica 11.

To make sure that our model predicting how two-photon transitions in non-centrosymmetric molecules are affected by shaped laser pulses is consistent with the experimental data, it is important to consider situations where the TPEF signal does not exhibit a square law dependence on laser intensity (i.e., the power dependence with $n = 2$) assumed in our considerations above. Such situations could be caused, for example, by saturation of the two-photon transition. Although saturation was not reached in our experiments, the deviations from a square law dependence of the TPE rate for molecules with large dipole moments have been suggested in ref 43. We simulate such deviations assuming that no interference occurs between the virtual and dipole pathways, i.e., the $2\operatorname{Re}\left([E^{(2)}(\omega)]^* \times aE^2(\omega/2)e^{i\phi}/(\omega/2)\right)$ terms obtained by expanding the integrands in the numerator and denominator of eq 7 are neglected. This assumption leads to the following modification of the formula for the TL-normalized TPEF signal:

$$\tilde{S}_{\text{shaped}}^{\text{TPEF}} \Big/ \tilde{S}_{\text{TL}}^{\text{TPEF}} = \frac{\int_{-\infty}^{\infty}\left[\left|E^{(2)}(\omega)\right|^2_{\text{shaped}} + \left|aE^2(\omega/2)e^{i\phi}/(\omega/2)\right|^2_{\text{shaped}}\right]d(\omega/2)}{\int_{-\infty}^{\infty}\left[\left|E^{(2)}(\omega)\right|^2_{\text{TL}} + \left|aE^2(\omega/2)e^{i\phi}/(\omega/2)\right|^2_{\text{TL}}\right]d(\omega/2)}. \tag{9}$$

We found that for this case, the value of the empirical constant $a$ in eq 9 that worked best was 0.18. As discussed in the Results and Discussion section, these simulations failed to simulate several features found in the experimental data, suggesting that the interference between virtual and dipole pathways is of paramount importance.

When simulating the TPEF signals and their ratios with SHG as functions of the width of the phase window, we found that the results obtained for broader windows agreed with the experimental data, but those obtained for the narrow ones did not. We considered the lack of agreement found for narrower phase windows to be caused by inhomogeneous broadening, which is quite significant in protic solutions of **FR0**-SB.[19] In doing so, we assumed that one might expect a reduction of the interference between the virtual and dipole pathways in TPE processes when the window widths become narrower. To simulate the effect of the postulated interference reduction in the TL-normalized TPEF signals as a result of inhomogeneous broadening, we formed a linear combination of eq 7, which assumes interference, and eq 9, in which interference is absent, and compared the resulting TPEF/SHG ratios with the corresponding experimental data. The relative contribution of eq 7 to the linear combination of eqs 7 and 9 was increased linearly from 0% for the 10 nm window width to 80% for the 50 nm window width. As shown in the Results and Discussion section, the model assuming the decrease of interference as a result of window width reduction is in much better agreement with the experimental data obtained for all phase windows.

## EXPERIMENTAL DETAILS

The experimental setup used a Ti:Sapphire oscillator (Coherent Vitara-S) producing pulses with a repetition rate of 80 MHz for excitation. The pulses were compressed and shaped with a pulse shaper (MIIPS Box 640, Biophotonic Solutions Inc.) using MIIPS.[33–42] The output beam, centered at 811 nm with a pulse duration of ~16 fs, was split with a 20:80 beam splitter (Thorlabs UFBS2080). The reflected beam (20% reflected) was frequency doubled in a β-BBO crystal to serve as our reference, which was detected with an Ocean Optics USB4000 spectrometer. The transmitted beam was focused by a 10 cm focal length convex lens onto a 1 cm cuvette containing the solutions of **FR0**-SB in methanol with a concentration of ~6 μM. The maximum laser power at the sample was ~50 mW with a peak intensity of ~$6\times10^8$ W/cm$^2$, which is at least one order of magnitude below the onset of saturation (cf. Figure S1 in the Supplementary Material of ref 19). TPEF was detected perpendicular to excitation with a 2.5 cm focal length convex lens to collimate the signal followed by a 10 cm focal length convex lens to focus the signal onto an Ocean Optics QE PRO spectrometer. The pulse chirp in both the reference signal and the fluorescence excitation beam paths were matched to a value of less than 50 fs$^2$ by the introduction of glass slides in the reference beam path so that the pulses in both paths were TL after pulse compression using MIIPS. All experiments were performed multiple times to ensure reproducibility of the data. The laser power dependence for TPE for the samples measured gave a nonlinear exponent of 1.90 ± 0.02. Deviation from the exact square law dependence is expected for molecules with significant dipole pathway contribution.[12,43]

## RESULTS AND DISCUSSION

Two-photon excitation of the super photobase **FR0**-SB leads to the electronically excited S$_1$ state **FR0**-SB*, which shows emission at ~21,000 cm$^{-1}$.[19] In protic solvents, excited-state proton transfer from the solvent to the photobase **FR0**-SB* leads to the formation of the excited protonated photobase **FR0**-HSB$^{+}$*, which emits at ~15,000 cm$^{-1}$.[18,19] Because the SHG spectrum provides an accurate measurement of $|E^{(2)}(\omega)|^2$, which, as shown by comparing Figures 1b and 1c, can be accurately modeled using eq 5, it serves in this study as a reference. In this study, we collected the frequency integrated TPEF signals using shaped and TL pulses for **FR0**-SB in methanol and compared them to the SHG reference data. Using the same phase functions as those utilized in Figure 1 and eq 5, we simulated the ratios of the frequency integrated TPEF signals obtained for shaped and TL pulses, defined by eq 7, and their SHG counterparts, defined by eq 8, using the window width of 40 nm and window phases varying from −2π to 2π. The results of these simulations are shown in Figure 3a. The corresponding experimental TPEF signals for **FR0**-SB in methanol and the experimental SHG reference spectra obtained using shaped pulses and normalized to their values under TL excitations are plotted in Figure 3b. The ratio of the fluorescence signal to the integrated laser signal amplitude reveals a maximum enhancement at ±0.75π and ±1.25π, when the probability for TPE exceeds the value predicted by $|E^{(2)}(\omega)|^2$ alone.

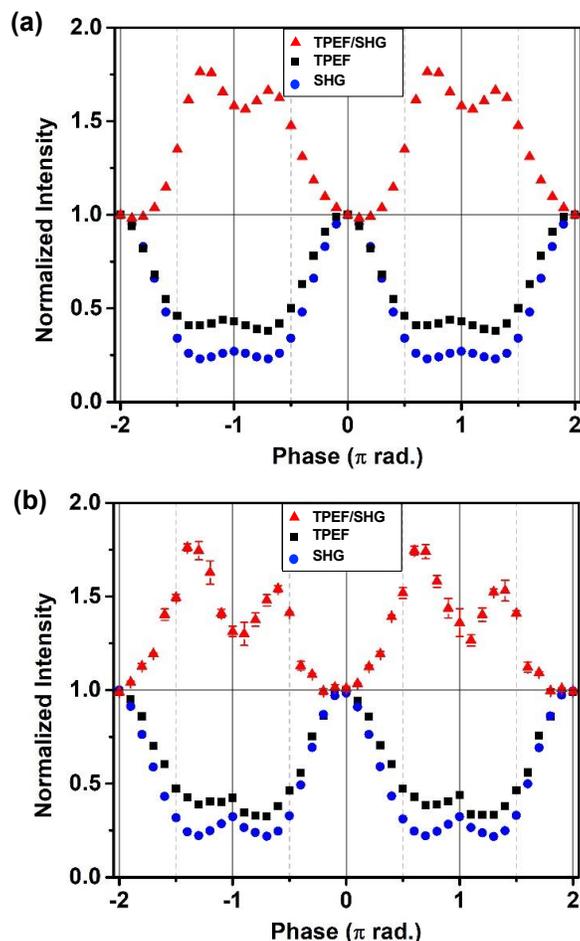

**Figure 3**. (a) Simulation of the TL-normalized TPEF signal based on eq 7 (black squares), the TL-normalized SHG signal based on eq 8 (blue circles), and the ratio of the two signals (red triangles) as functions of the spectral phase. The simulation parameters are given in the Theory section. (b) Experimental data corresponding to (a) obtained following TPE of **FR0**-SB in methanol using 16.6 fs pulses and a 40 nm window as a function of the spectral phase. The TL-normalized TPEF signal is shown in black squares, the TL-normalized SHG signal is shown in blue circles, and the ratio between the two signals is shown in red triangles.

The excellent agreement between the simulation and experimental data shown in Figure 3 demonstrates that the anticipated control of the quantum interference between the virtual and dipole pathways that contribute to TPE in large non-centrosymmetric molecules in solution is indeed possible. To examine the possibility of controlling the interference between the two pathways, we performed a series of additional measurements in which the TPEF signals for **FR0**-SB in methanol and the corresponding SHG reference signals were obtained for several different phase window widths ranging from 10 nm to 50 nm and plotted as functions of the window phase scanned between −2π and 2π. The results of these additional measurements are shown in Figure 4.

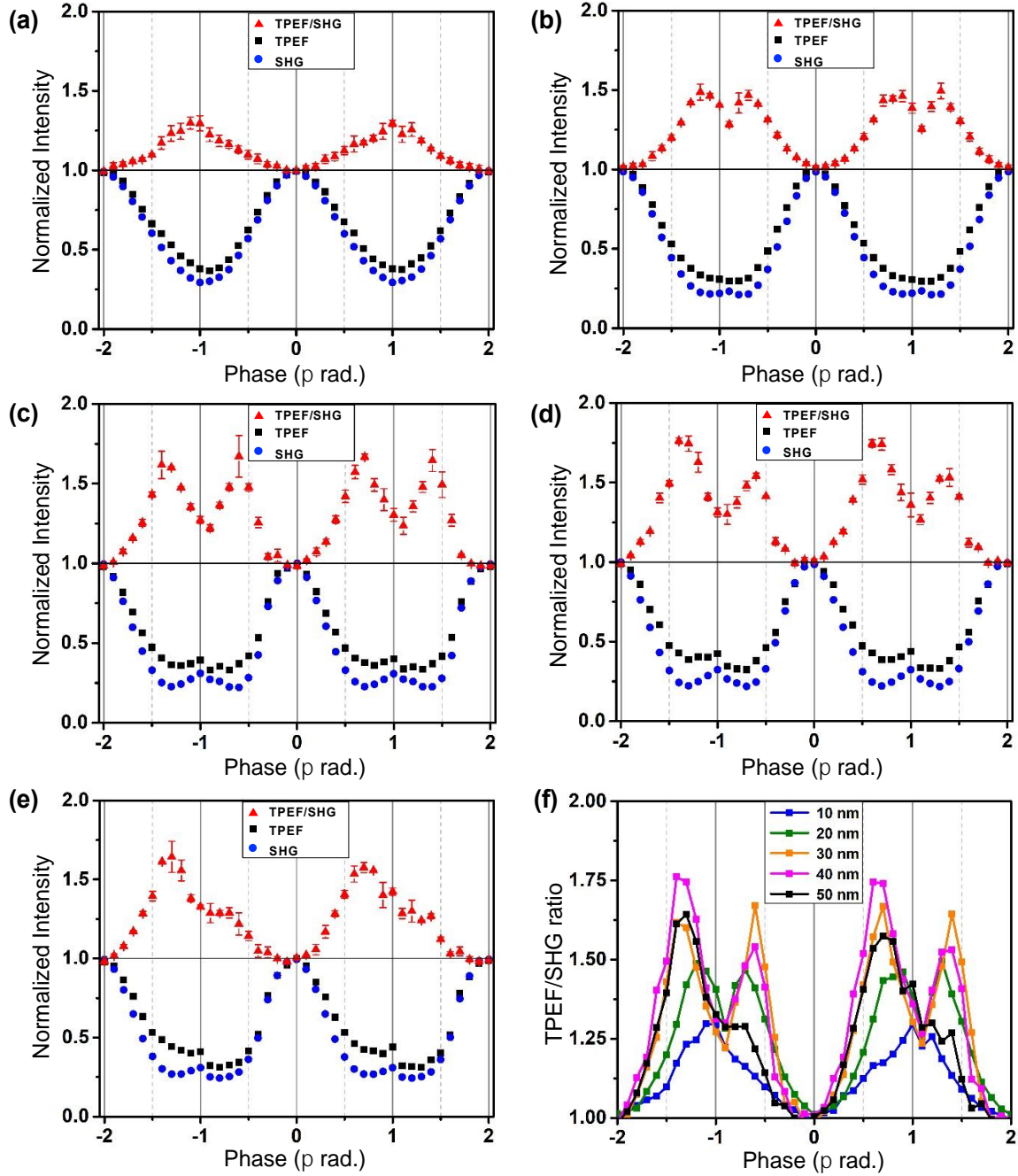

**Figure 4**. Experimental results for the frequency integrated TPEF signal for **FR0**-SB in methanol (black squares) and the reference SHG signal of the laser (blue circles), along with the TPEF/SHG signal ratios with error bars (red triangles), plotted as functions of the spectral phase for window widths of (a) 10 nm, (b) 20 nm, (c) 30 nm, (d) 40 nm, and (e) 50 nm. The TPEF/SHG signal ratios for all window sizes examined in (a)–(e) are shown in (f).

The results in Figure 4 show a clear dependence of the TPEF/SHG signal ratios on the window width. When the window is 10 nm wide, the maximum enhancement in the normalized TPEF is only 25% above that of the normalized SHG signal. However, for a window width of 40 nm we observe that TPEF can be up to 75% greater than the SHG signal. As the window width increases further to 50 nm, the observed enhancement diminishes. The optimum window width is found to be about 2/3 of the FWHM of the laser pulse spectrum (cf. Figure 1a).

As already mentioned in the Theory section and as implied by comparison of Figures 3a and 3b, the parameter $a$ measuring the relative contribution of the dipole and virtual pathways and the phase lag $\phi$ characterizing the dipole pathway can be fit to closely match the experimental results shown in Figure 3b. However, eq 7 fails to reproduce the experimental data for some of the window widths examined in Figure 4. This is demonstrated in Figure 5, where we show the simulated TPEF and SHG signals, along with the corresponding TPEF/SHG ratios, as functions of the window phase scanned between $-2\pi$ and $2\pi$ for window widths ranging from 10 nm to 50 nm, using the same parameters as used in Figure 3a and described in the Theory section. While the agreement between experiment and simulation is best for the larger phase window widths, the simulation results for the 10 and 20 nm windows based on eq 7 do not match the corresponding experimental data.

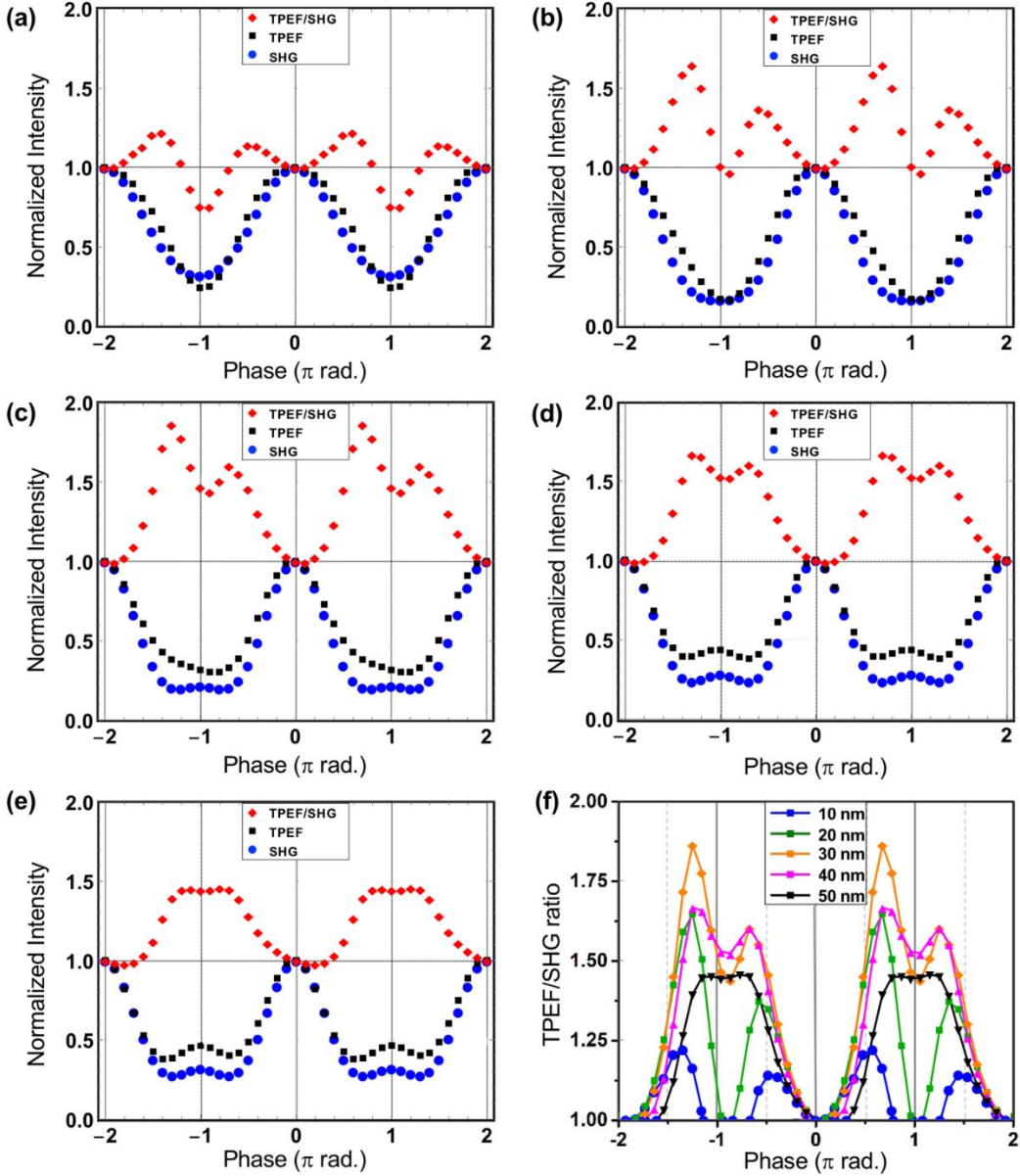

**Figure 5**. Simulation results for the frequency integrated TPEF signal for **FR0**-SB in methanol (black squares) and the reference SHG signal of the laser (blue circles), along with the TPEF/SHG signal ratios (red triangles), based on eqs 7 and 8, plotted as functions of the spectral phase for window widths of (a) 10 nm, (b) 20 nm, (c) 30 nm, (d) 40 nm, and (e) 50 nm. The simulated TPEF/SHG signal ratios for all window sizes examined in (a)–(e) are shown in (f). The simulation parameters are given in the Theory section.

To understand the disagreement between simulations based on eq 7 and experiment for narrower window widths, we explored two alternative explanations. In the first explanation, we assumed that there is no interference between the virtual and dipole pathways and that the presence of the dipole pathway attenuates the laser power dependence of two-photon transitions. This idea of reduced power dependence was explored by Meath who discussed the intensity dependence of two-photon absorption due to the direct permanent dipole moment excitation mechanism.[43] By employing the rotating wave approximation, Meath showed that the TPE rate due to direct permanent dipole mechanism is not necessarily proportional to the square of the laser intensity and may display a lower power dependence, which for very low intensities can be a linear function of the laser intensity. We simulated the TPEF signals assuming that no interference occurs between the virtual and dipole pathways using eq 9. We found that for these simulations the value of the empirical constant $a$ in eq 9 that worked best was 0.18. Figure 6a shows the results of the simulations based on eq 9, presented as the ratio between the calculated, TL-normalized, TPEF (eq 9) and SHG (eq 8) signals. The results in Figure 6a demonstrate that the neglect of the interference between the virtual and dipole pathways fails to describe the experimental TPEF/SHG ratios shown in Figure 4f. In particular, the theoretical TPEF/SHG ratios simulated in this manner fail to reproduce the dip in the experimental TPEF/SHG ratios, which, as shown in Figure 4f, should occur when the phase is $\pm\pi$.

In the second explanation of the failures of eq 7 for the narrower window widths, we assumed that eq 7 is capable of modeling the interference, but we also postulated that the interference at smaller window widths may weaken due to inhomogeneous broadening. This could be rationalized by the fact that **FR0**-SB in methanol represents a situation where a large molecule is hydrogen-bonded to a strongly polar solvent,[19] so that the inhomogeneous broadening might dominate the TPE line width(s) when dealing with narrow phase windows. The simulations shown in Figure 5, which do not take into account inhomogeneous broadening, show a narrow but significant interference when the amplitude of the phase window is $\pm\pi$, especially for the 10 and 20 nm windows. However, the experimental data shown in Figure 4 indicate that inhomogeneous broadening significantly dampens this interference and a broad background of 'unshaped' field observed for windows narrower than 30 nm becomes dominant. The effect of inhomogeneous broadening was taken into account by the simulations shown in Figure 6b, where the contribution from eq 7 to the linear combination of eq 7, which assumes interference, and eq 9, in which the interference is absent, was linearly increased from 0% for the 10 nm window width to 80% for the 50 nm window width. These modified simulations are in much better agreement with the experimentally observed TPEF/SHG ratios, shown in Figure 4f and reproduced in Figure 6c, than those obtained by dividing eq 7 by eq 8 or those presented in Figure 6a that ignore interference.

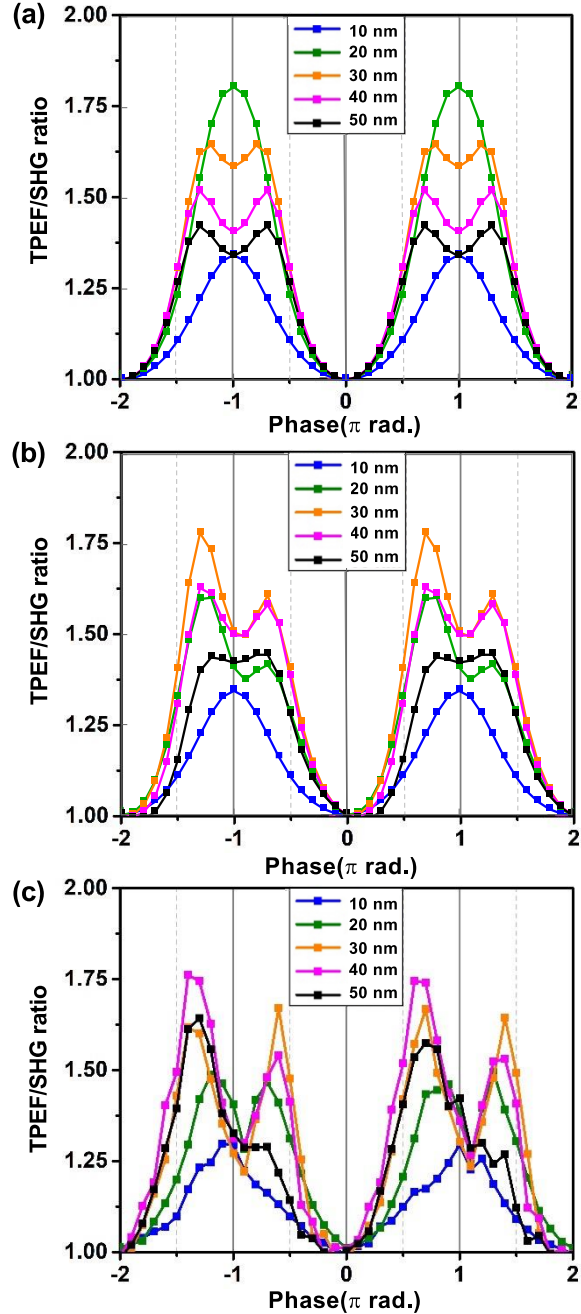

**Figure 6**. Calculated [panels (a) and (b)] and experimentally determined [panel (c)] TPEF/SHG ratios as functions of phase values for different window widths. (a) Results obtained assuming that the interference between the virtual and dipole pathways in the TPEF signal can be neglected, as in eq 9. (b) Results obtained assuming that the TPEF signal can be simulated using a linear combination of eq 7, which assumes interference, and eq 9, where interference is absent, with the contribution from eq 7 being increased linearly with the window width (see text for details). (c) Experimental data, as shown in Figure 4f, reproduced here to facilitate comparison.

## CONCLUSIONS

The dependence of TPEF of **FR0**-SB in methanol on the spectral phase of femtosecond broadband pulses, which reflects on the quantum interference between the dipole and virtual pathways that contribute to TPE, was examined experimentally and theoretically. When normalized for pulse duration, enhancements by a factor of up to 1.75 in the experimentally observed TPEF were found. Simulations taking into account the different responses of the virtual and dipole pathways to external fields and the effect of pulse shaping on two-photon transitions, were found to be in good agreement with the experimental measurements, but we encountered situations corresponding to narrow spectral window widths for which there were disagreements between theory and experiment. We explained these disagreements by postulating that the interference between the virtual and dipole pathways at smaller window widths weakens as a result of inhomogeneous broadening, showing that the model assuming the decrease of interference as a result of window width reduction is in very good agreement with the experimentally observed TPEF/SHG signal ratios obtained for all phase windows. At the same time, we demonstrated that simulations ignoring the interference between the virtual and dipole pathways and considering an attenuated power dependence only fail to reproduce the experimental data. Findings from the present study, in addition to their potential impact on applications that depend on TPE, may be relevant to ongoing research on the two-photon transitions induced by entangled photons,[44–49] where even molecules with weaker permanent dipole moments may exhibit strong quantum effects, such as transparencies.

## ACKNOWLEDGMENTS

The collaboration between synthesis, theory, and experiments for the understanding and development of super photoreagents for precision chemistry is funded by a seed grant from DARPA and AMRDEC (W31P4Q-20-1-0001). Partial support comes from NSF (Grant No. CHE1836498 to MD), NIH (Grant Nos. 2R01EY016077-08A1 and 5R01EY025383-02 R01 to GJB, and R01GM101353 to BB), and the U.S. DOE (Grant No. DE-FG02-01ER15228 to PP). The views and conclusions contained in this document are those of the authors and should not be interpreted as representing the official policies, either expressed or implied, of the Defense Advanced Research Projects Agency, the U.S. Army, or the U.S. Government.

## DATA AND AVAILABILITY

The data that support the findings of this study are available within the article.

## REFERENCES

(1) Meshulach, D.; Silberberg, Y. Coherent Quantum Control of Two-Photon Transitions by a Femtosecond Laser Pulse. *Nature* **1998**, *396*, 239–242.

(2) Walowicz, K. A.; Pastirk, I.; Lozovoy, V. V.; Dantus, M. Multiphoton Intrapulse Interference. 1. Control of Multiphoton Processes in Condensed Phases. *J. Phys. Chem. A* **2002**, *106*, 9369–9373.

(3) Lozovoy, V. V.; Pastirk, I.; Walowicz, K. A.; Dantus, M. Multiphoton Intrapulse Interference. II. Control of Two- and Three-Photon Laser Induced Fluorescence with Shaped Pulses. *J. Chem. Phys.* **2003**, *118*, 3187–3196.

(4) Lozovoy, V. V.; Dantus, M. Systematic Control of Nonlinear Optical Processes Using Optimally Shaped Femtosecond Pulses. *ChemPhysChem* **2005**, *6*, 1970–2000.

(5) Pastirk, I.; Dela Cruz, J. M.; Walowicz, K. A.; Lozovoy, V. V.; Dantus, M. Selective Two-Photon Microscopy with Shaped Femtosecond Pulses. *Opt. Express* **2003**, *11*, 1695–1701.

(6) Lozovoy, V. V.; Xu, B.; Shane, J. C.; Dantus, M. Selective Nonlinear Optical Excitation with Pulses Shaped by Pseudorandom Galois Fields. *Phys. Rev. A* **2006**, *74*, 041805.

(7) Schelhas, L. T.; Shane, J. C.; Dantus, M. Advantages of Ultrashort Phase-Shaped Pulses for Selective Two-Photon Activation and Biomedical Imaging. *Nanomedicine* **2006**, *2*, 177–181.

(8) Ogilvie, J. P.; Débarre, D.; Solinas, X.; Martin, J.-L.; Beaurepaire, E.; Joffre, M. Use of Coherent Control for Selective Two-Photon Fluorescence Microscopy in Live Organisms. *Opt. Express* **2006**, *14*, 759–766.

(9) Isobe, K.; Suda, A.; Tanaka, M.; Kannari, F.; Kawano, H.; Mizuno, H.; Miyawaki, A.; Midorikawa, K. Multifarious Control of Two-Photon Excitation of Multiple Fluorophores Achieved by Phase Modulation of Ultra-Broadband Laser Pulses. *Opt. Express* **2009**, *17*, 13737–13746.

(10) Labroille, G.; Pillai, R. S.; Solinas, X.; Boudoux, C.; Olivier, N.; Beaurepaire, E.; Joffre, M. Dispersion-Based Pulse Shaping for Multiplexed Two-Photon Fluorescence Microscopy. *Opt. Lett.* **2010**, *35*, 3444–3446.

(11) Jagatap, B. N.; Meath, W. J. On the Competition between Permanent Dipole and Virtual State Two-Photon Excitation Mechanisms, and Two-Photon Optical Excitation Pathways, in Molecular Excitation. *Chem. Phys. Lett.* **1996**, *258*, 293–300.

(12) Kondo, A. E.; Meath, W. J. Two-Color Multiphoton Transitions in Molecular Beam Electric Resonance Studies: Rotating Wave versus Floquet, and On- versus Off- Resonance, Calculations. *J. Chem. Phys.* **1996**, *104*, 8312–8320.

(13) Dudovich, N.; Dayan, B.; Gallagher Faeder, S. M.; Silberberg, Y. Transform-Limited Pulses Are Not Optimal for Resonant Multiphoton Transitions. *Phys. Rev. Lett.* **2001**, *86*, 47–50.

(14) Drobizhev, M.; Meng, F.; Rebane, A.; Stepanenko, Y.; Nickel, E.; Spangler, C. W. Strong Two-Photon Absorption in New Asymmetrically Substituted Porphyrins: Interference between Charge-Transfer and Intermediate-Resonance Pathways. *J. Phys. Chem. B* **2006**, *110*, 9802–9814.

(15) Alam, M. M.; Chattopadhyaya, M.; Chakrabarti, S. On the Origin of Large Two-Photon Activity of DANS Molecule. *J. Phys. Chem. A* **2012**, *116*, 11034–11040.

(16) Sheng, W.; Nairat, M.; Pawlaczyk, P. D.; Mroczka, E.; Farris, B.; Pines, E.; Geiger, J. H.; Borhan, B.; Dantus, M. Ultrafast Dynamics of a "Super" Photobase. *Angew. Chem. Int. Ed.* **2018**, *57*, 14742–14746.

(17) Lahiri, J.; Moemeni, M.; Kline, J.; Borhan, B.; Magoulas, I.; Yuwono, S. H.; Piecuch, P.; Jackson, J. E.; Dantus, M.; Blanchard, G. J. Proton Abstraction Mediates Interactions between the Super Photobase FR0-SB and Surrounding Alcohol Solvent. *J. Phys. Chem. B*

**2019**, *123*, 8448–8456.

(18) Lahiri, J.; Moemeni, M.; Magoulas, I.; Yuwono, S. H.; Kline, J.; Borhan, B.; Piecuch, P.; Jackson, J. E.; Blanchard, G. J.; Dantus, M. Steric Effects in Light-Induced Solvent Proton Abstraction. *Phys. Chem. Chem. Phys.* **2020**, *22*, 19613–19622.

(19) Lahiri, J.; Moemeni, M.; Kline, J.; Magoulas, I.; Yuwono, S. H.; Laboe, M.; Shen, J.; Borhan, B.; Piecuch, P.; Jackson, et al. Isoenergetic Two-Photon Excitation Enhances Solvent-to-Solute Excited-State Proton Transfer. *J. Chem. Phys.* **2020**, *153*, 224301.

(20) Göppert-Mayer, M. Über Elementarakte Mit Zwei Quantensprüngen. *Ann. Phys.* **1931**, *401*, 273–294.

(21) Honig, B.; Jortner, J.; Szöke, A. Theoretical Studies of Two-Photon Absorption Processes. I. Molecular Benzene. *J. Chem. Phys.* **1967**, *46*, 2714–2727.

(22) Webman, I.; Jortner, J. Energy Dependence of Two-Photon-Absorption Cross Sections in Anthracene. *J. Chem. Phys.* **1969**, *50*, 2706–2716.

(23) Dick, B.; Hohlneicher, G. Importance of Initial and Final States as Intermediate States in Two-photon Spectroscopy of Polar Molecules. *J. Chem. Phys.* **1982**, *76*, 5755–5760.

(24) Meath, W. J.; Power, E. A. On the Importance of Permanent Moments in Multiphoton Absorption Using Perturbation Theory. *J. Phys. B At. Mol. Phys.* **1984**, *17*, 763–781.

(25) Shapiro, M.; Brumer, P. Laser Control of Product Quantum State Populations in Unimolecular Reactions. *J. Chem. Phys.* **1986**, *84*, 4103–4104.

(26) Zhu, L.; Kleiman, V.; Li, X.; Lu, S. P.; Trentelman, K.; Gordon, R. J. Coherent Laser Control of the Product Distribution Obtained in the Photoexcitation of HI. *Science* **1995**, *270*, 77–80.

(27) Chen, Z.; Brumer, P.; Shapiro, M. Coherent Radiative Control of Molecular Photodissociation via Resonant Two-Photon versus Two-Photon Interference. *Chem. Phys. Lett.* **1992**, *198*, 498–504.

(28) Chen, Z.; Brumer, P.; Shapiro, M. Multiproduct Coherent Control of Photodissociation via Two-Photon versus Two-Photon Interference. *J. Chem. Phys.* **1993**, *98*, 6843–6852.

(29) Meshulach, D.; Silberberg, Y. Coherent Quantum Control of Multiphoton Transitions by Shaped Ultrashort Optical Pulses. *Phys. Rev. A* **1999**, *60*, 1287–1292.

(30) Präkelt, A.; Wollenhaupt, M.; Sarpe-Tudoran, C.; Baumert, T. Phase Control of a Two-Photon Transition with Shaped Femtosecond Laser-Pulse Sequences. *Phys. Rev. A* **2004**, *70*, 063407.

(31) Weiner, A. M. Femtosecond Pulse Shaping Using Spatial Light Modulators. *Rev. Sci. Instrum.* **2000**, *71*, 1929–1960.

(32) Silberberg, Y. Quantum Coherent Control for Nonlinear Spectroscopy and Microscopy. *Annu. Rev. Phys. Chem.* **2009**, *60*, 277–292.

(33) Lozovoy, V. V.; Pastirk, I.; Dantus, M. Multiphoton Intrapulse Interference. IV. Ultrashort Laser Pulse Spectral Phase Characterization and Compensation. *Opt. Lett.* **2004**, *29*, 775–777.

(34) Xu, B.; Gunn, J. M.; Dela Cruz, J. M.; Lozovoy, V. V.; Dantus, M. Quantitative Investigation of the Multiphoton Intrapulse Interference Phase Scan Method for Simultaneous Phase Measurement and Compensation of Femtosecond Laser Pulses. *J. Opt. Soc. Am. B* **2006**, *23*, 750–759.

(35) Pastirk, I.; Zhu, X.; Martin, R. M.; Dantus, M. Remote Characterization and Dispersion Compensation of Amplified Shaped Femtosecond Pulses Using MIIPS. *Opt. Express* **2006**, *14*, 8885–8889.

(36) Xu, B.; Coello, Y.; Lozovoy, V. V.; Harris, D. A.; Dantus, M. Pulse Shaping of Octave Spanning Femtosecond Laser Pulses. *Opt. Express* **2006**, *14*, 10939–10944.

(37) Harris, D. A.; Shane, J. C.; Lozovoy, V. V.; Dantus, M. Automated Phase Characterization and Adaptive Pulse Compression Using Multiphoton Intrapulse Interference Phase Scan in Air. *Opt. Express* **2007**, *15*, 1932–1938.

(38) Dantus, M.; Lozovoy, V. V.; Pastirk, I. MIIPS Characterizes and Corrects Femtosecond Pulses. *Laser Focus World* **2007**, *43*, 101–104.

(39) Zhu, X.; Gunaratne, T. C.; Lozovoy, V. V.; Dantus, M. In-Situ Femtosecond Laser Pulse

Characterization and Compression during Micromachining. *Opt. Express* **2007**, *15*, 16061–16066.

(40) Lozovoy, V. V.; Xu, B.; Coello, Y.; Dantus, M. Direct Measurement of Spectral Phase for Ultrashort Laser Pulses. *Opt. Express* **2008**, *16*, 592–597.

(41) Coello, Y.; Lozovoy, V. V.; Gunaratne, T. C.; Xu, B.; Borukhovich, I.; Tseng, C.; Weinacht, T.; Dantus, M. Interference without an Interferometer: A Different Approach to Measuring, Compressing, and Shaping Ultrashort Laser Pulses. *J. Opt. Soc. Am. B* **2008**, *25*, A140–A150.

(42) Pestov, D.; Lozovoy, V. V.; Dantus, M. Single-Beam Shaper-Based Pulse Characterization and Compression Using MIIPS Sonogram. *Opt. Lett.* **2010**, *35*, 1422–1424.

(43) Meath, W. J. On the Optimization, and the Intensity Dependence, of the Excitation Rate for the Absorption of Two-Photons Due to the Direct Permanent Dipole Moment Excitation Mechanism. *AIP Adv.* **2016**, *6*, 075202.

(44) Saleh, B. E. A.; Jost, B. M.; Fei, H.-B.; Teich, M. C. Entangled-Photon Virtual-State Spectroscopy. *Phys. Rev. Lett.* **1998**, *80*, 3483–3486.

(45) Lee, D.-I.; Goodson III, T. Entangled Photon Absorption in an Organic Porphyrin Dendrimer. *J. Phys. Chem. B* **2006**, *110*, 25582–25585.

(46) Harpham, M. R.; Süzer, Ö.; Ma, C.-Q.; Bäuerle, P.; Goodson III, T. Thiophene Dendrimers as Entangled Photon Sensor Materials. *J. Am. Chem. Soc.* **2009**, *131*, 973–979.

(47) Guzman, A. R.; Harpham, M. R.; Süzer, Ö.; Haley, M. M.; Goodson III, T. G. Spatial Control of Entangled Two-Photon Absorption with Organic Chromophores. *J. Am. Chem. Soc.* **2010**, *132*, 7840–7841.

(48) Upton, L.; Harpham, M.; Suzer, O.; Richter, M.; Mukamel, S.; Goodson III, T. Optically Excited Entangled States in Organic Molecules Illuminate the Dark. *J. Phys. Chem. Lett.* **2013**, *4*, 2046–2052.

(49) Burdick, R. K.; Varnavski, O.; Molina, A.; Upton, L.; Zimmerman, P.; Goodson III, T. Predicting and Controlling Entangled Two-Photon Absorption in Diatomic Molecules. *J. Phys. Chem. A* **2018**, *122*, 8198–8212.